\documentclass[twocolumn]{aastex631}

\usepackage{amsmath, amsfonts}
\newcommand{\redtext}[1]{{#1}}
\DeclareMathOperator\erf{erf}

\newcommand{\kms}{$\rm km\,s^{-1}$}

\begin{document}

\title{Direct Statistical Constraints on the Natal Kick velocity of a Black Hole in an X-ray Quiet Binary }

\author[0000-0001-7852-7484]{Sharan Banagiri}
\email{sharan.banagiri@northwestern.edu}
\affiliation{Center for Interdisciplinary Exploration and Research in Astrophysics (CIERA), Northwestern University, 1800 Sherman Ave, Evanston, IL 60201, USA }

\author[0000-0002-2077-4914]{Zoheyr Doctor}
\affiliation{Center for Interdisciplinary Exploration and Research in Astrophysics (CIERA), Northwestern University, 1800 Sherman Ave, Evanston, IL 60201, USA }

\author[0000-0001-9236-5469]{Vicky Kalogera}
\affiliation{Center for Interdisciplinary Exploration and Research in Astrophysics (CIERA), Northwestern University, 1800 Sherman Ave, Evanston, IL 60201, USA }

\author[0000-0001-9879-6884]{Chase Kimball}
\affiliation{Center for Interdisciplinary Exploration and Research in Astrophysics (CIERA), Northwestern University, 1800 Sherman Ave, Evanston, IL 60201, USA }

\author[0000-0001-5261-3923]{Jeff J. Andrews}
\affiliation{Department of Physics, University of Florida, 2001 Museum Rd, Gainesville, FL 32611, USA}



\begin{abstract}

\noindent
 In recent years, a handful of ``dark" binaries have been discovered with a non-luminous compact object. Astrometry and radial velocity measurements of the bright companion allow us to measure the post-supernova orbital elements of such a binary. In this paper, we develop a statistical formalism to use such measurements to infer the pre-supernova orbital elements, and the natal kick imparted by the supernova (SN). We apply this formalism to the recent discovery of an X-ray quiet binary with a black hole, VFTS 243, in the Large Magellanic Cloud. Assuming an isotropic, Maxwellian distribution on natal kicks and using broad agnostic mass priors, we find that kick velocity can be constrained to \redtext{$V_k < 72$~\kms}~and the dispersion of the kick distribution to \redtext{$\sigma_k < 68  $ \kms}~at 90 \% confidence. \redtext{We find that a Blaauw kick cannot be ruled out and }that at least about $0.6 M_{\odot}$ was lost during the supernova with 90 \% confidence. The pre-SN orbital separation is found to be robustly constrained to be around \redtext{$0.3$ AU}.

\end{abstract}

\keywords{Black holes, Binary stars, Bayesian statistics}


\section{Introduction} \label{sec:intro}
\noindent Galactic binaries with a compact object companion provide one of the few ways to study the population of black holes (BHs) in our galaxy. The list of known stellar-mass BHs in our galactic environment is small. About 23 BHs in accreting binary systems luminous in X-rays have been discovered with mass and spin measurements~\citep{Corral-Santana:2015fud, Reynolds:2020jwt}. These have been complemented in recent years by a handful of ``dark'' binaries, where the non-luminous companion's mass is constrained by measurement of orbital parameters through astrometric or spectroscopic means~\citep{2018MNRAS475L15G, 2019A&A632A3G, Shenar:2022tmt, El-Badry:2022, 2022arXiv221005003C}. In addition, a free-floating compact object has been recently detected~\citep{OGLE:2022gdj, 2022ApJ933L23L} through lensing, although its mass and thereby its nature are somewhat in dispute. 

Recent years have also seen the population of extra-galactic stellar-mass black holes (BHs) come into focus driven primarily by the gravitational-wave (GW) discovery of binary black hole (BBH) mergers by the Laser Interferometer Gravitational-Wave Observatory (LIGO) and Virgo detectors~\citep{LIGOScientific:2016aoc, LIGOScientific:2020ibl, LIGOScientific:2021djp, LIGOScientific:2021usb}. Understanding the multiple evolutionary pathways of BH formation has become an important astrophysical problem. For instance, GWs and X-ray binary observations are starting to show us the mass and spin distributions of BHs which contain signatures of their evolutionary pathways (for e.g.~\cite{LIGOScientific:2020kqk, LIGOScientific:2021psn, Corral-Santana:2015fud, Qin:2018sxk, Reynolds:2020jwt, Draghis:2022ngm}). 

 While the coming years promise many BH detech tions through GWs, these systems are formed at higher redshifts and lower metallicities and are therefore not representative of the Galactic environment. Moreover, it is possible that X-ray binaries and BBHs discovered through GWs represent different, somewhat distinct parts of the population of BHs in binaries~\citep{Fishbach:2021xqi, Gallegos-Garcia:2022rve, Liotine:2022vwq}. Gaia however is expected to uncover the galactic population of binaries containing a compact object through precision astrometric measurements of object (see for e.g.~\cite{Andrews:2019, Breivik:2017cmy, Janssens:2022, Chawla2022}).

The binary's orbital dynamics can be impacted by SN physics, in particular through any kicks imparted to the compact object during its birth. In general, these kicks are caused by the recoil on the remnants from the spatially asymmetric momentum loss during the SN explosion, e.g.~\citet{Janka:2012wk, Wongwathanarat:2012zp}. Such natal kicks can change the eccentricity and the post-SN orbital separation, and perhaps most importantly can unbind binaries if binaries are wide and kicks are large. Even in the limit of negligible SN asymmetry in which case the remnant does not receive a recoil kick, the orbital parameters and the center-of-mass velocity of the binary could still change due to mass loss. This is sometimes called a Blaauw kick~\citep{Blaauw:1961BAN15265B}. Through their impact on the post-SN separation, the kicks also affect the time delay between star formation and binary coalescence through Peters's formula~\citep{Peters:1964, OShaughnessy:2009szr, Mapelli:2017hqk, Fishbach:2021mhp}. Natal kicks, particularly in the case of a second SN, can also affect the spin-tilts in the field binary formation channel~\citep{Kalogera:1996rm, Farr:2011gs}. 


While the natal kick velocities for NSs are reasonably well understood~\citep{Lyne:1994, Hobbs:2005yx, Kapil:2022blf}, these constraints mainly come from analyses of the proper motions of galactic pulsars. Under a Maxwellian model, the dispersion of the natal kick velocities for galactic neutron stars was estimated to be $\sigma = 265 $ \kms. On the other hand, the magnitude and nature of the natal kicks for BH remnants have remained much more uncertain. Since proper motion measurements are much harder for unbound BHs (however, see ~\cite{Andrews:2022cwi}), similar constraints are not readily obtainable. Some studies have used measurements of proper motion, masses and orbital parameters of the X-ray binaries to trace back their evolutionary history, obtaining their natal kick history in the process~\citep{Willems:2004kk, Fragos:2008hg, Wong:2011eg, Wong:2013vya}. In particular, recently~\cite{Kimball:2022xbp} studied the low-mass X-ray binary MAXI J1305-704, finding that its BH received a natal kick of at least 70 
\kms~ with 95\% confidence. A similar lower limit of 80 \kms~ was obtained for XTE J1118 + 480 by~\cite{Fragos:2008hg}. 

In this paper, we show how the measurements of the post-SN orbital parameters of a binary with a BH/NS and a luminous star can be used to constrain natal kick velocity and pre-SN orbital parameters. We develop a Bayesian statistical formalism that can be generally applied as long as the pre-SN eccentricity is negligible and the binary is wide with negligible interactions.

We apply this method to the recent observation by~\citet{Shenar:2022tmt} of VFTS 243, an X-ray quiet BH in a binary in the Large Magellanic Cloud (LMC). The luminous companion is reported to be a $\simeq 25 M_{\sun}$ O-type star while the mass of the black hole is reported to be $ \simeq 10$ $M_{\sun}$. A non-degenerate dim alternative to the BH is excluded at high statistical confidence. Through radial velocity measurements from the Fibre Large Array Multi Element Spectrograph (FLAMES)~\citep{2011A&A530A108E}, orbital parameters of this system were inferred. In particular, the orbital period and eccentricity are tightly constrained at $10.4031$ days and $0.017$, respectively. Observations of the system using Chandra detected no signs of X-ray luminosity, indicating a quiescent BH with little accretion from its companion over its history. 

\cite{Stevance:2022cqo} recently analyzed VFTS 243 using the BPASS population synthesis code to study the evolutionary history and progenitors of the system. Comparing the results of their population synthesis models with the post-SN properties of VFTS 243, they find that the SN had a low explosion energy and low recoil kick velocities of less than 33 \kms~at 90\% confidence. In this paper, we treat this as an inverse problem and develop a Bayesian statistical formalism for inferring the natal kick velocity directly from observations. Thereby, we develop a way to statistically constrain certain progenitor properties, without using population synthesis. 

The rest of this paper is organized as follows. First, in Sec.~\ref{Sec:VFTS243} we briefly describe the measurements of the masses and the orbital parameters of VFTS 243,  and its properties which make it a good candidate for the kind of analysis done here. In Sec.~\ref{Sec:Dynamics} we briefly describe the dynamics of natal kicks and how they impact the orbital parameters. We develop the statistical formalism for inferring natal kicks and pre-SN orbital parameters in Sec.~\ref{Sec:Stats}. In Sec.~\ref{Sec:priors} we describe priors and selection effects, followed by the results of this formalism when applied to VFTS 243 in Sec.~\ref{Sec:results}. We discuss some implications of this work in Sec.~\ref{Sec:implications} followed by a summary in Sec.~\ref{Sec:conclusion}.

\section{Properties of VFTS 243}
\label{Sec:VFTS243}
\noindent 
\cite{Shenar:2022tmt} report the discovery of an X-ray quiet dark binary VFTS 243 in the Tarantula nebula of the Large Magellanic Cloud. The system was analyzed using spectra obtained from the FLAMES spectrograph of the European Southern Observatory. The primary star in the binary is an O-type star with a mass of about $25 \pm 2.5 M_{\odot}  $ with the mass of the dark companion estimated to be around $10.1 \pm 2.0 M_{\odot}$. The minimum mass of the companion is constrained to be at least $8.7 M_{\odot}$. The values of some important parameters are given in Table ~\ref{Tab:postSN}. Full posterior distributions of these parameters are shown in the Supplementary Materials of \citet{Shenar:2022tmt}. Through spectral analysis and comparing the data with a mock data set, they rule out a faint non-degenerate companion concluding the companion is a degenerate star. The minimum mass limit implies that it cannot be a neutron star and has to be a BH. 

\cite{Shenar:2022tmt} also rule out X-ray emission from the binary, and thereby any significant accretion, through upper limits from Chandra on the X-ray luminosity, $\log L_X < 32.84~\rm erg . s^{-1}$. The primary was observed to be rapidly rotating with a period that is not synchronized with the orbital period. Thereby they conclude that tidal effects and accretion after the SN can be ignored, and the orbital parameters after the SN have been maintained. They also point out that the rapid rotation of the primary and the presence of CNO-processed material in the spectrum indicates a period of accretion before the SN, likely when the BH-progenitor was passing through a giant phase. This strongly implies that the pre-SN orbit was circularized.

\begin{table}[t!]
\centering
\begin{tabular}{|c | c|} 
 \hline
 Parameter & Value  \\ [0.5ex] 
 \hline
 $P_f$ & $10.4031^{+0.0004}_{-0.0004} \, \rm days$ \\ 
 \hline
 $e_f$ & $0.017^{+0.012}_{-0.012}$  \\
 \hline
 $M_1$ & $25.0^{+2.3}_{-2.3} \, M_{\sun}$  \\
 \hline
 $M_2$  & $10.1^{+2.0}_{- 2.0} \, M_{\sun}$  \\
 \hline
 $M_{\rm tot}$  & $36.3^{+3.8}_{-5.5} \, M_{\sun}$ \\ 
 \hline
 $\Gamma_f$ & $260.2 \pm 0.9$ \kms \\
 \hline
\end{tabular}
\caption{Table of measured vales of post-SN orbital parameters~\citep{Shenar:2022tmt}. The uncertainties correspond to $1\sigma~(68 \%)$ error bars.}
\label{Tab:postSN}
\end{table}

\section{Dynamics of a natal kick in a binary system}
\label{Sec:Dynamics}

\begin{figure*} [ht]
    \centering
    \includegraphics[width=0.8 \textwidth]{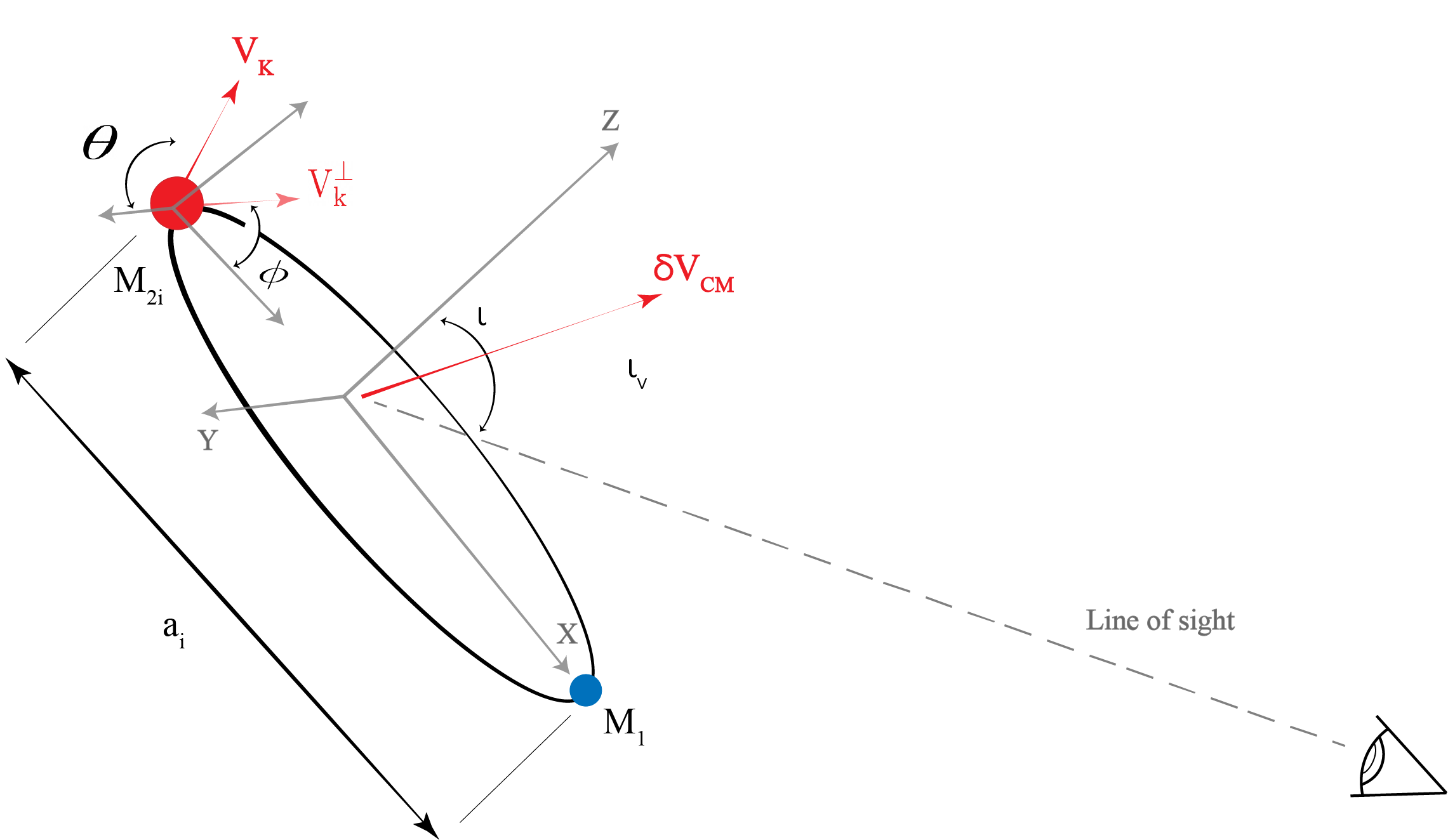}
    \caption{Schematic of the pre-SN geometry of the binary system, along with the coordinate system being used. Also shown are the kick velocity $V_k$, the angle $\theta$, the CM kick $\delta V_{CM}$, and the angle the latter makes with the line-of-sight $\iota_v$. Note that $\iota_v$ is different from the pre-SN inclination angle $\iota$.}
    \label{Fig:Orbital_schematic}
\end{figure*}

\noindent
We use as the starting point the derivation from \citet{Kalogera:1996rm} that relates how the post-SN orbital parameters are related to the pre-SN parameters after a natal kick. A schematic of the kick along with all the parameters is provided in Fig.~\ref{Fig:Orbital_schematic}. Some of the relevant details of the derivation are reproduced below. 

Firstly, following~\citet{Kalogera:1996rm} we define the parameters

\begin{equation}
         \alpha =  \frac{a_f}{a_i} \qquad \text{ and } \qquad \beta =  \frac{M_1 + M_{2f}}{M_1 + M_{2i}}, 
\end{equation}
where $a_i~\text{and } a_f$ are the pre-SN and post-SN orbital separations respectively. Throughout this paper, we label the star that undergoes the SN as `2' and its (originally less massive) companion as `1'. Therefore, $M_{2i}~\text{and} M_{2f} $ represent the pre-SN and the post-SN mass, respectively. $M_1$ is the mass of the other binary companion which we assume doesn't change appreciably during this process. We also define the pre-SN orbital velocity, 

\begin{equation}
    V_r = \sqrt{\frac{G (M_1 + M_{2i})}{ a_i}},
\end{equation}
where $G$ is the Newtonian gravitational constant. We will further assume that the pre-SN orbit is circular. For VFTS 243, this follows the argument made in~\cite{Shenar:2022tmt} that the rapid rotation and signs of pre-SN accretion onto the primary from the BH progenitor show that the orbit was circularized by mass transfer. 

Now, let the SN remnant receive a natal kick of magnitude $V_k$ in the pre-SN reference frame of star 2. Suppose the kick makes an angle $\theta$ with the Y-axis and subtends an azimuthal angle $\phi$ in the X-Z plane as shown in Fig.~\ref{Fig:Orbital_schematic}. Then using a spherical coordinate system aligned with the pre-SN orbital velocity, the pre-SN, and post-SN orbital parameters can be related as~\citep{Kalogera:1996rm, Andrews:2019ome}, 

\begin{equation}
    \alpha = \frac{\beta}{2 \beta - v^2_k - 1 - 2 v_k \cos \theta},
\label{Eq:calc_p}
\end{equation}

\begin{equation}
    1 - e^2 = \frac{1}{\alpha \beta} \left( 2 \beta - \frac{\beta}{\alpha} - v^2_k \sin^2 \theta \cos^2 \phi \right),
\label{Eq:calc_e}
\end{equation}
where $v_k = V_k/ V_r$. The center-of-mass (CM) of the binary also receives a recoil kick $\vec{V}_{\rm CM}$ with a magnitude given by~\citep{Kalogera:1996rm, Andrews:2019ome},

\begin{equation}
    v_{\rm CM}^2 = \kappa_1 + \kappa_2 \frac{(2\alpha -1)}{\alpha} - \kappa_3 \frac{v_k \cos \theta + 1}{\sqrt{\beta}},
    \label{Eq:calc_vcm}
\end{equation}
where $v_{\rm CM} = V_{\rm CM} / V_k $ and $\kappa_1, \kappa_2$ and $\kappa_3$ are: 

\begin{equation}
    \begin{split}
        \kappa_1 = & \; \frac{M_{2i}^2}{(M_1 + M_{2i})^2}, \\
        \kappa_2 = & \; \frac{M_{2f}^2}{(M_1 + M_{2i})(M_1 + M_{2f})},\\
        \kappa_3 = & \; \sqrt{4 \kappa_1 \kappa_2}.
    \end{split}
\end{equation}
Note that through Eqs.~\ref{Eq:calc_p},~\ref{Eq:calc_e} and~\ref{Eq:calc_vcm}, the post-SN parameters deterministically depend on the pre-SN parameters if all the masses are known.

\section{Statistical formalism for measuring natal kicks}

\label{Sec:Stats}
\noindent
This section will set up a general statistical formalism to infer kick velocities from the orbital parameters of a dark binary. We assume that the post-SN orbital and mass parameters of the system are known, in particular the eccentricity $e_f$, the orbital period $P_f$ (from which the post-SN separation $a_f$ can be calculated) and the masses $M_1, M_{2f}$. We shall also assume that the post-SN radial CM velocity, $\Gamma_f$, has been measured, although we make no assumptions about constraints on inclination. 

Let $ \pi (P_f, e_f, M_1, M_{2f}, \Gamma_f)$ represent the priors used for the post-SN parameters. Using Bayes theorem we relate the posteriors and priors on the post-SN parameters to the likelihood over the radial-velocity data $d$,

\begin{equation}
\begin{split}
        P(P_f, e_f, M_1, M_{2f} \, | \, d) \,  \propto \, &   \mathcal{L}(d \, | \, P_f, e_f, M_1, M_{2f} ) \\ & \times \pi (P_f, e_f, M_1, M_{2f}).
\end{split}
\end{equation}

Our goal is to estimate the pre-SN parameters, in particular, $a_i$ and the natal kick magnitude $V_k$. Using Bayes theorem, we can once again relate likelihoods and posteriors on these parameters

\begin{equation}
    \begin{split}
    P(a_i, V_k, M_{2i}, M_1 | \, d)  \propto & \, \mathcal{L}(d \, | \, a_i, V_k, M_{2i}, M_1) \\
    & \times \pi ( a_i, V_k, M_{2i}, M_1 ).
    \end{split}
\end{equation}
We define a hyper-prior $\pi( P_f, e_f, \Gamma_f, M_{2f} | a_i, V_k, M_{2i} )$ that describes the distribution of post-SN parameters for a given set of pre-SN parameters. Then we can write,

\begin{equation}
    \begin{split}
   P(a_i, V_k, M_{2i} |  d)  \propto & \, \pi (  a_i, V_k, M_{2i} ) \int  \, dP_f \, de_f \, d \Gamma_f \, dM_1 \, dM_{2f} \\ \times &  \,  \mathcal{L}(d | P_f, e_f, \Gamma_f, M_1, M_{2f} ) \\ \times & \,\pi( P_f, e_f, \Gamma_f, M_{2f} | a_i, V_k, M_{2i} ) \pi(M_1).
    \end{split}
    \label{Eq:hyper-bayes}
\end{equation}
We will also assume that the hyper-prior can be factorised so that

\begin{equation}
\begin{split}
    \pi( P_f, e_f, \Gamma_f, M_{2f} | a_i, V_k,  M_{2i}) & =  \pi( P_f, e_f, \Gamma_f | a_i, V_k, M_{2i}) \\& \times  \pi(M_{2f} | M_{2i} ).
\end{split}
\end{equation}

This factorization assumes that mass loss from the SN is only dependent on the pre-SN mass and not on the kick velocity or orbital period. This may not be strictly true in nature, but we take it as a starting assumption here, which can be built upon in future work.

In order to derive $\pi( P_f, e_f, \Gamma_f | a_i, V_k, M_{2i})$,  we first look at the simplified case where only the orbital parameters are measured, and the radial CM velocity is completely unconstrained. Under the assumption that the natal kicks have no particular directional preference in the pre-SN rest frame of star 2 and marginalizing over those angular variables, an analytical probability distribution for $\alpha, e$ can be derived~\citep{Kalogera:1999dx, Andrews:2019ome}:

\begin{equation}
    \pi (\alpha, e_f | \beta, v_k) = \frac{e_f}{\pi \alpha 
    v_k} \sqrt{\frac{\beta^3}{C_1 C_2}}. 
\end{equation}
\label{Eq:P_alpha_e}
The terms $C_1$ and $C_2$ are functions of $\alpha, e_f$:

\begin{eqnarray}
    C_1 & = & 2 - \frac{1}{\alpha} - \alpha (1 - e^2),  \\  C_2 & = & 4 v^2_k - \left(2 \beta - \frac{\beta}{\alpha} - v^2_k - 1 \right)^2 - 4 \beta C_1. 
\end{eqnarray}

We now extend this and generalize to the case where the radial CM velocity is measured as well. We do this by additionally marginalizing over the unknown pre-SN CM velocity $\Gamma$. This is assumed to be drawn from the environmental distribution of radial velocities, which we model by a normal distribution, 

\begin{equation}
    p(\Gamma) = \frac{1}{\sqrt{2 \pi \sigma^2_0}} \exp \left [ - \frac{(\Gamma - \mu_0 )^2}{2\sigma_0^2} \right].
    \label{Eq:vel_prior}
\end{equation}
Also marginalizing over pre-SN inclination with a uniform prior~\footnote{This is necessary because we only care about the radial CM velocity in the particular case of VFTS 243. But it can be straightforwardly extended to include the proper motion as well}, we get  (see Appendix.~\ref{Sec:Appdx1} for derivation)

\begin{equation}
\small
\begin{split}
   & \pi (\alpha, e_f, \Gamma_f  | \beta, V_k) = \frac{ e_f \beta^{3/2} }{ (2 \pi)^{5/2} \sigma_0 \alpha v_k \sqrt{C_1 C_2  }} \\ & \times \left \{ \erf \left(\frac{\Gamma_f + \delta V_{CM} - \mu_0}{\sqrt{2 \sigma_0^2}} \right) - \erf \left(\frac{\Gamma_f - \delta V_{CM} - \mu_0}{\sqrt{2 \sigma_0^2}} \right)   \right \}, 
   \label{Eq:VCM_incl_prob}
\end{split}
\end{equation}

Since we desire a distribution on $P_f$ rather than $\alpha$, we perform an additional Jacobian transformation $\left |\frac{\partial \alpha}{\partial P_f} \right|$. Using Kepler's third law, 

\begin{equation}
  \alpha \propto  a_f \propto P_f^{2/3} \implies  \frac{1}{\alpha} \left| \frac{\partial \alpha}{\partial P_f} \right| = \frac{2}{3} P_f^{-1}.
\end{equation}
Substituting this in Eq.~\ref{Eq:VCM_incl_prob} we get, 

\begin{equation}
\small
\begin{split}
   & \pi (P_f, e_f, \Gamma_f  | \beta, V_k) = \frac{ 2 e_f \beta^{3/2} }{ (2 \pi)^{5/2} 3 \sigma_0 P_f v_k \sqrt{C_1 C_2  }} \\ & \times \left \{ \erf \left(\frac{\Gamma_f + \delta V_{CM} - \mu_0}{\sqrt{2 \sigma_0^2}} \right) - \erf \left(\frac{\Gamma_f - \delta V_{CM} - \mu_0}{\sqrt{2 \sigma_0^2}} \right)   \right \}.
\end{split}
\end{equation}

Armed with a mass prior (the choices of which are discussed in more detail in Sec.~\ref{Sec:mass_prior}) we can put together the hyperprior $\pi( P_f, e_f, \Gamma_f, M_{2f} | a_i, V_k,  M_{2i})$ and calculate posteriors on $a_i, V_k~\text{and } M_{2i}$.  If we already have fair draws from $ P(e, a_f, M_1, M_{2f} \, | \, d) $, we can approximate this expression as

\begin{equation}
\begin{split}
 &  P(a_i, V_k, M_{2i}  |  d)  \propto  \pi (  a_i, V_k, M_i^{\rm tot} ) \\ & \times \sum_n\, \frac{ \pi( P^n_f, e^n_f, M^n_{2f} | a_i, V_k, \beta, M_{2i} ) }{\pi (e_f^n, P_f^n, M_{2f}^n) }.
\end{split}
\label{Eq:resampling_final_eq}
\end{equation}
where $n$ is an index over the Monte Carlo samples of post-SN parameters. \redtext{The prior on $M_1$ is assumed to be unchanged in the hierarchical step and hence cancels out in this expression}. 

\noindent There are two bounds of validity for Eq.~\ref{Eq:P_alpha_e} that have to be considered. Firstly, from the argument that the post-SN orbit has to contain the position of the star just before the SN,~\citet{1975A&A3961F} showed that $(1 + e )^{-1} < \alpha < (1 - e)^{-1}$. This is equivalent to the condition that $C_1 > 0$. Furthermore, since the probability in Eq.~\ref{Eq:P_alpha_e} has to be real, we also need $C_2 > 0$. It can be shown that this implies~\citep{Andrews:2019ome}

\begin{equation}
    e^2 < 1 - \frac{\left( \beta + \alpha (v^2_k - 2 \beta -1)\right)^2}{4 \beta \alpha^3}.
\end{equation}

\subsection{Assumptions}
\noindent
We list here, explicitly, the assumptions that went into this derivation. 

\begin{enumerate}
  \item The orbital eccentricity before the SN was negligible. This follows from the argument that mass transfer circularized the pre-SN orbit. See Sec.~\ref{Sec:VFTS243} and~\citet{Shenar:2022tmt}. 
  
  \item The natal kick is isotropic; i.e. it has no apriori directional preference. 

  \item $M_1$ has not changed since the SN happened. This follows from the assumption that there has been no mass transfer between the primary and the BH. 
   
  \item Orbital parameters from immediately after the SN have been preserved and tidal synchronization can be neglected. These assumptions again follow~\cite{Shenar:2022tmt}'s argument that there is no evidence of post-SN mass transfer or tidal synchronization.
  
  \item Again following \cite{Shenar:2022tmt}, the inclination of the system is assumed to be unconstrained. In principle, however, it would be straightforward to also draw from an inclination posterior distribution if it was better constrained.

\end{enumerate}

\section{Priors}
\label{Sec:priors}
\noindent
\cite{Shenar:2022tmt} use the following priors for the post-SN parameters; $P_f \sim \rm U(9.9, 10.9) \, days, \,  e \sim U(0, 0.2)$,  and $\Gamma_f \sim U(240, 280)$ \kms, while $M_{2f}$ is drawn from a flat prior with a mass function of 0.581. $M_1$ was assumed to be drawn from a Gaussian $M_1 \sim \mathcal{N}(25.02 M_{\sun},  2.32 M_{\sun})$\footnote{Private communications with Tomer Shenar and Leonardo Almeida.}.

For the pre-SN parameters, we adopt a uniform prior on the pre-SN separation; {$a_i \sim \rm U(0.01, 2)$ AU}. For the natal kick magnitude we have two prior models. The first is a \textsc{uniform} prior on the kick magnitude {$V_k \sim \rm (0, 200)$ \kms} and isotropic in direction in the pre-SN frame. The second prior draws the kick from a $3$D Gaussian which implies a \textsc{Maxwellian} prior on the velocity magnitude when marginalized over the kick directions 

\begin{equation}
    \pi \left (V_k | \sigma \right) = \frac{V^2_k}{\sigma^3_k} \sqrt{\frac{2}{\pi}} \exp \left( - \frac{V^2_k}{2 \sigma^2_k} \right), 
    \label{Eq:Maxwell_prior}
\end{equation}
with the kick-velocity dispersion $\sigma_k$ treated as an independent unknown variable, directly estimable from the data\footnote{An easy way to draw from this is through the cumulative distribution, \begin{equation}
    \text{CDF}(V_k) = \text{erf} \left( \frac{V_k}{\sqrt{2} \sigma_k}\right) - \frac{V_k}{\sigma_k} \sqrt{\frac{2}{\pi}} \exp \left( - \frac{V_k^2}{2 \sigma^2_k} \right).  \end{equation}} . We further use a uniform prior on the dispersion, $\sigma_k \sim \rm U(0.01, 200)$ \kms~to match the bounds for the velocity prior in the uniform model. 

Note that the velocity priors technically decouple the mass lost during the SN from the recoil kick, which is not entirely realistic. There are SN kick recipes that allow for some dependence between progenitor masses and kick velocity~\citep{Bray:2016mab, Mandel:2020qwb}, however they are based on phenomenological fits to simulations due to the paucity of observational data. Since we consider only one system here, the simple independent models we use is assumed to be sufficient. We leave to future work the consideration of a coupled prior that combines mass loss and kick velocities.

\subsection{Mass priors}
\label{Sec:mass_prior}
\noindent
We explore two mass prior choices. The first is a broadly agnostic mass prior, where we let {$M_{2i} \sim \rm U(5, 30) M_{\sun}$} and $\pi (M_{2f} | M_{2i})$ also to be uniform but bounded by $M_{2i}$, i.e. {$M_{2f}  \sim \rm U(5 M_{\sun} , M_{2i})$}. The second kind of prior is fixed-$\beta$, i.e. a delta function in $\beta$. To convert this to a prior on $M_{2i}$, we first compute the Jacobian

\begin{equation}
    \pi_{\beta}(M_{2i}) =   \pi(\beta) \left| \frac{d \beta}{d M_{2i}} \right |.
\end{equation}

Setting $\pi (\beta)$ to be a delta function, we get

\begin{equation}
    \pi_{\beta}(M_{2i}) = \frac{\beta^2}{(M_1 + M_{2f})}.
\end{equation}

\subsubsection{Selection effects}

\noindent When measuring $\sigma_k$, we also have to account for selection effects arising from the fact that large natal kicks - more likely to occur when the underlying dispersion $\sigma_k$ is large - can unbind a system. In other words, in the presence of large $\sigma_k$ we are more likely to see binaries which suffered comparatively smaller natal kicks. To account for these selection effects, we follow the statistical prescription from ~\citet{Mandel:2018mve}. We first define $p_{\rm survival} (\alpha, e, V)$ as the fraction of systems that survive the kick, marginalized over the kick direction~\citep{Andrews:2019ome}.

\begin{equation}
    P_{\rm surv} (v_k) = \frac{2 \beta - (v_k - 1)^2}{4v_k}.
    \label{Eq:Surv_prob}
\end{equation}

The selection function is then just the probability that a system survives the natal kick, marginalized over the distribution of pre-SN parameters i.e.;

\begin{equation}
\begin{split}
\alpha_{\rm surv}(M_{2i}, a_i, \sigma_k) = \int & \, dV_k \, d M_1 \, d \beta \, p_{\rm surv} (V) \, p(V_k | \sigma_k) \\ & \times \pi (\beta | M_1, M_{2i}) \, P (M_1),
\end{split}
\label{Eq:selecfxn}
\end{equation}
where $p(V_k  | \sigma)$ is given by Eq.~\ref{Eq:Maxwell_prior}, and $\pi(\beta | M_1, M_{2i})$ is the distribution of $\beta$ under the mass priors described in Sec.~\ref{Sec:priors}.

We estimate the integral Eq.~\ref{Eq:selecfxn} stochastically, drawing from the prior probabilities of the parameters. Once drawn, we then correct for the systematic effects as, 

\begin{equation}
    P(a_i, V_k, M_{2i} | d) = \frac{ P\left (a_i, V_k, M_{2i} | d \right)}{\alpha_{\rm surv}(M_{2i}, a_i, \sigma_k)}
\end{equation}

\begin{figure*}[ht]
    \centering
    \includegraphics[width=0.9 \textwidth]{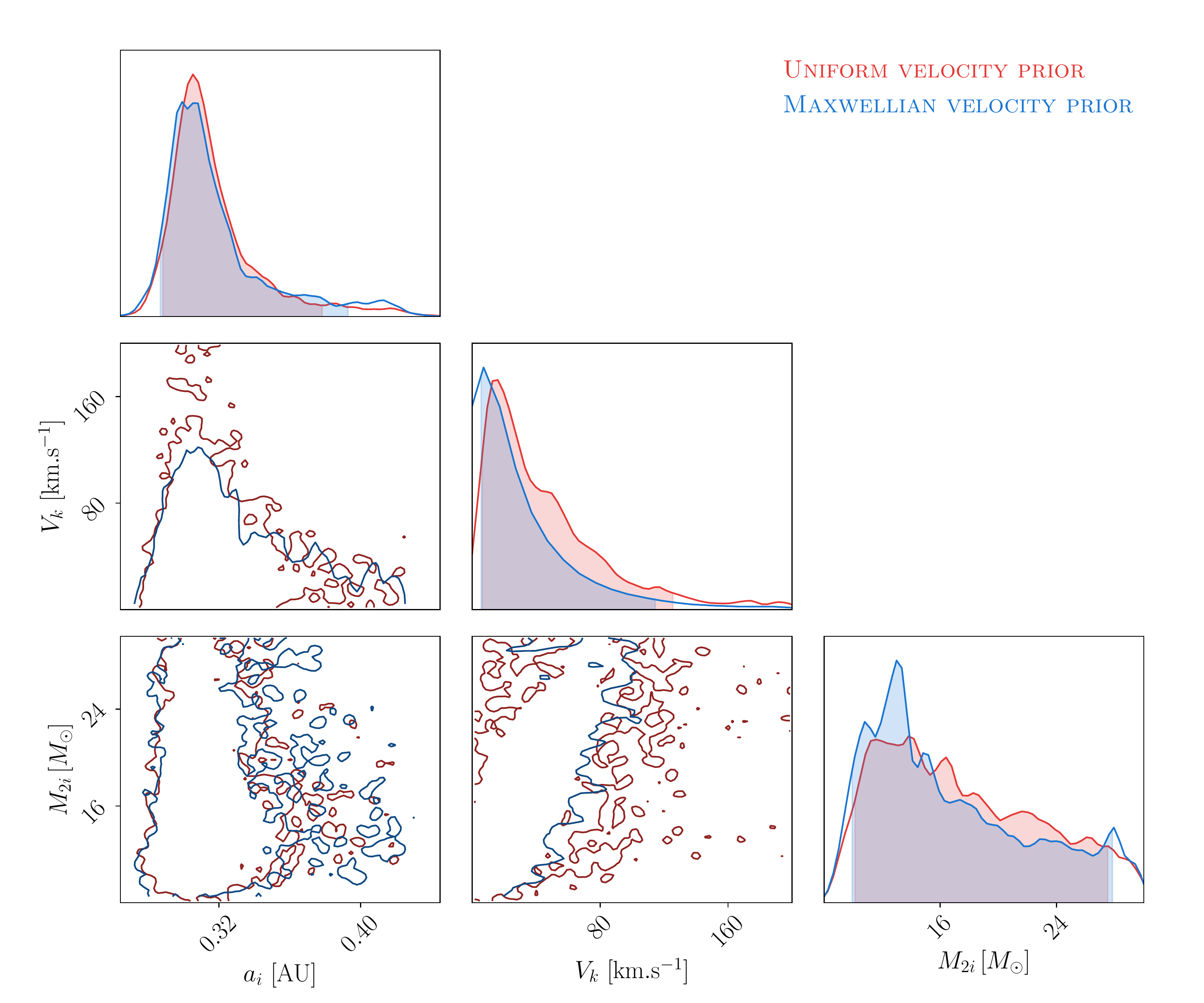}
    \caption{Posterior probability distributions of pre-SN separation, natal  kick velocity and the pre-SN mass for the Maxwellian and uniform velocity models. The shaded regions correspond to 90\% credible intervals. }
    \label{Fig:main_corner_plot}
\end{figure*}
\subsection{Bounds of validity}

\section{Results and discussions}
\label{Sec:results}

\noindent
We now apply this formalism to the results from~\cite{Shenar:2022tmt} for VFTS 243. We use the nested sampler \textsc{dynesty}~\citep{Skilling:2006, Dynesty:speagle} for sampling over the pre-SN parameters using Eq.~\ref{Eq:resampling_final_eq}. 

Figure~\ref{Fig:main_corner_plot} shows the constraints on kick velocity and pre-SN separation for the case of uninformative and \redtext{broad} uniform mass priors. While we do not measure the kick velocity well, it is constrained to less \redtext{than $72$ \kms~and $97$} \kms~at $90\%$ confidence for the Maxwellian and the uniform velocity priors, respectively. \redtext{In both cases, the kick velocity posteriors are fully consistent with $V_k = 0$}. We, therefore, conclude that the kick velocity was likely small and is not compatible with the much higher kick velocities observed for neutron star remnants~\citep{Hobbs:2005yx}. This is in line with naive physical expectations of relatively lower mass loss during BH formation compared to a neutron star, though larger kicks are in principle possible \citep{Janka2013MNRAS, Fragos:2008hg, Kimball:2022xbp}. Note that while the Maxwellian limits are smaller it has a much longer tail that extends up to $\sim 500$ \kms. This is likely due to the selection effects which make higher velocity dispersion harder to rule out.

\begin{figure}[ht]
    \centering
    \includegraphics[width=0.48 \textwidth]{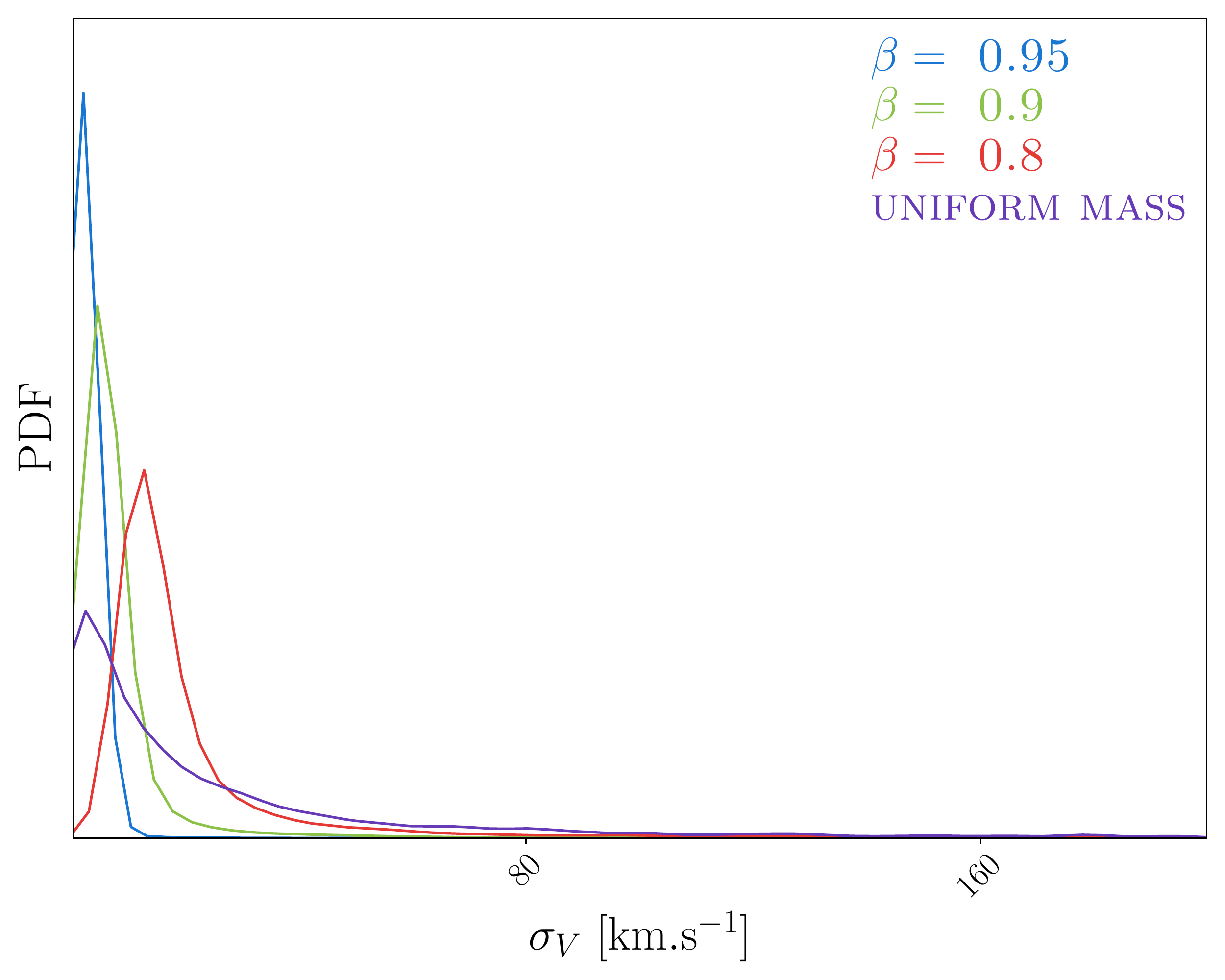}
    \caption{Posterior probability distributions of the kick-velocity dispersion under the Maxwellian model for different mass priors.}
    \label{Fig:Maxwell_disperion}
\end{figure}

Assuming that this system is representative of the BH population, we now check what this means for the distribution of kick velocities in the Maxwelian model. Fig.~\ref{Fig:Maxwell_disperion} plots the posterior distribution of the dispersion, $\sigma_k$, for a uniform mass prior and two different fixed-$\beta$ priors. Note that $\beta = 0.95, 0.9 \text{ and } 0.8$ corresponds to $M_{2i} \simeq 12 M_{\sun}, 14 M_{\sun}$ and $\simeq 19 M_{\sun}$ respectively. While the tails of these distributions go past $150$ \kms, there is little support past $\simeq 80$ \kms\ implying again that, if there is a common natal velocity dispersion for all BH-generating supernovae, it is likely small. For instance, the uniform mass prior yields a $90 \%$ upper limit of \redtext{$68$ \kms} for $\sigma_k$. 

Both kick models give nearly similar estimates for the mass lost, $\Delta M = M_{2i} - M_{2f}$ during the SN. With the uniform mass prior, the Maxwellian model gives a median value of \redtext{$\Delta M = 3.6^{+10.6}_{-3.2} M_{\odot}$}, while the uniform velocity model gives \redtext{$\Delta M = 4.0^{+10.4}_{-3.7} M_{\odot}$}. The $90 \%$ lower limits on the mass lost during the SN are \redtext{$0.69 M_{\odot}$} and \redtext{$0.77 M_{\odot}$} respectively for the two models.  

\subsection{Pre-SN orbital separation}

\begin{figure}[b]
    \centering
    \includegraphics[width=0.48 \textwidth]{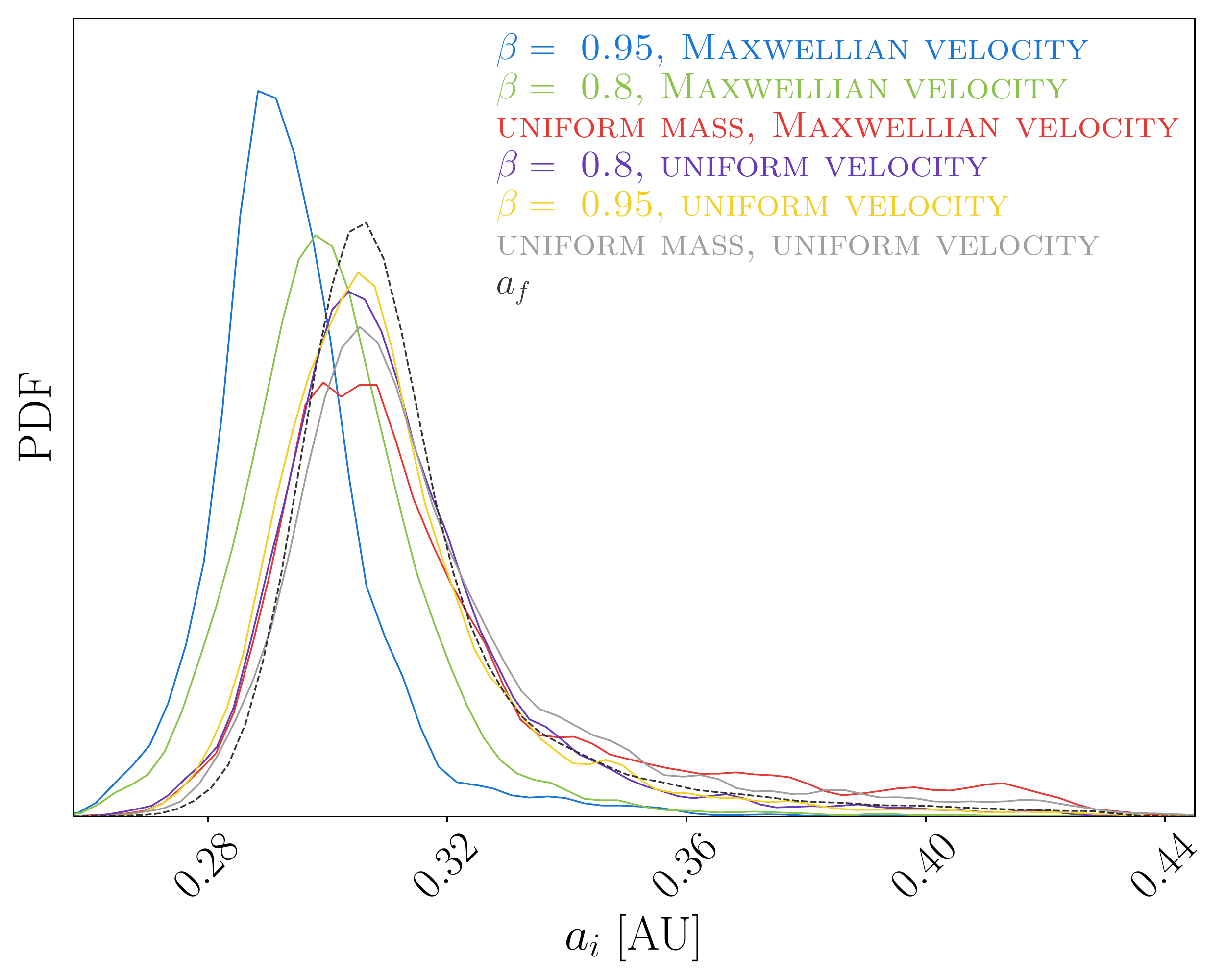}
    \caption{Posterior probability distribution of the pre-SN separation for different mass and velocity distributions. We see that the final posterior is quite insensitive to prior assumptions. For reference, the dashed line shows the post-SN semi-major axis $a_f$.}
    \label{fig:ai_plot}
\end{figure}

\noindent We see that the pre-SN orbital separation is constrained quite well at \redtext{$a_i \simeq 0.3 $} AU in both the uniform and Maxwellian velocity models in Fig.~\ref{Fig:main_corner_plot}. We further explore any dependency of this on mass models in Fig.~\ref{fig:ai_plot} by including several fixed-$\beta$ priors. The constraint on $a_i$ is seen to be quite robust and not strongly affected by prior assumptions on mass or velocity. \redtext{For reference, we also plot in Fig.~\ref{fig:ai_plot} the post-SN separation. The pre-SN and post-SN orbital are consistent with what we would expect with symmetric mass loss (see Sec.~\ref{Sec:Blaauw})}.

\subsection{Blaauw Kick}
\label{Sec:Blaauw}
\noindent

Under the limit that the natal kick velocity $V_K \rightarrow 0$, the binary receives only a CM kick from mass loss that is sometimes called the Blaauw kick~\citep{Blaauw:1961BAN15265B}. While the Maxwellian model rules out only a Blaauw kick for VFTS 243, this is somewhat artificial since the prior $\pi (V_k = 0 | \sigma) = 0$, (see Eq.~\ref{Eq:Maxwell_prior}). On the other hand,  $V_k = 0$ is entirely consistent with the posterior for the kick velocity in the uniform velocity model, as seen in Fig.~\ref{Fig:main_corner_plot}. This implies that a Blaauw kick can't be ruled out for VFTS 243. If the SN indeed imparted no recoil kick, this places a strong constraint on the mass lost of $\Delta M \simeq 0.60 M_{\odot}$~\citep{Hills:1983}, consistent with the lower limits estimated in Sec.~\ref{Sec:results}. 

\redtext{The symmetric mass loss in a Blaauw kick can also cause a change in binary orbital parameters. In particular~\citep{Blaauw:1961BAN15265B}}, 

\begin{equation}
    \begin{split}
    \alpha = & \frac{\beta}{2 \beta - 1} \\
    e_f = & \frac{1 - \beta}{\beta}
    \end{split}
    \label{Eq:Blaauw}
\end{equation}
\redtext{Combining these two expressions, we get $a_i = a_f(1-e_f) \simeq 0.3 AU$, which is consistent with the posteriors from Fig.~\ref{Fig:main_corner_plot}, further lending credence to low kick velocities}. 

\redtext{The structure of the covariance between the posteriors of $M_{2i}$ and $V_k$ in Fig.~\ref{Fig:main_corner_plot} can also be explained by considering the interplay between the effects of simple mass loss and the recoil from the natal kick. At higher $M_{2i}$, i.e. lower $\beta$ values, mass loss itself can cause a large change in orbital parameters as seen from Eq.~\ref{Eq:Blaauw}. In particular, as $\beta$ approaches $0.5$, the post-SN eccentricity induced by mass loss can become quite high. A higher natal recoil would then be needed as a counterbalance to maintain the small eccentricity value that is physically observed as seen in Fig.~\ref{Fig:main_corner_plot}.}

\subsection{Scaled-Hobbs Kick}
\noindent
We further explore a scaled Hobbs distribution for BH natal kick velocities. In this proposed distribution the dispersion of natal kicks for BH-generating supernovae is given by

\begin{equation}
    \sigma_{k, \rm BH} = \frac{M_{\rm NS}}{M_{\rm BH}} \times \sigma_{k, \rm  NS}
\end{equation}
where we use a fiducial neutron star mass of $M_{\rm NS} = 1.4 M_{\odot}$. Adopting $\sigma_{k, \rm NS} = 265$ \kms\ from~\cite{Hobbs:2005yx} and $M_{\rm BH} \simeq 10 M_{\sun}$, we perform a run fixing $\sigma_{\rm BH} = 37.1$ \kms~. Contrasting this number with Fig.~\ref{Fig:Maxwell_disperion}, we see it is clearly in the bulk of the $\sigma_k$ posterior and not ruled out. Indeed, unlike~\cite{Stevance:2022cqo}, we find that the scaled-Hobbs prior is not heavily disfavored compared to a uniform velocity draw with the Bayes factor between the two models being \redtext{$\mathcal{B}^{\rm Hobbs}_{\rm uniform} = 1.37$. However, it is disfavored compared to the Maxwellian model where $\sigma_k$ is allowed to vary with $\mathcal{B}^{\rm Maxwell}_{\rm Hobbs} = 4.41$}~\footnote{we assume the same prior probabilities for all three models}. While these specific numbers depend on several modeling choices and are only based on one one system, they do imply that the scaled-Hobbs model of BH natal cannot be ruled out for VFTS 243.

\section{Implications}
\label{Sec:implications}

Our analysis of VFTS 243 broadly demonstrates that the natal kick received by the BH in this system was small; and in particular much smaller than those received by neutron stars. This is consistent with the population synthesis analysis done by~\cite{Stevance:2022cqo} for this system using the BPASS code, wherein they find a natal kick velocity of the black hole of $<33$ \kms\ at the 90\% credible level. We both find low kick values and find that a Blaauw kick cannot be ruled out. However, we do find that a scaled-Hobbs prior cannot be ruled out.   

\begin{figure}[t]
    \centering
    \includegraphics[width=0.5 \textwidth]{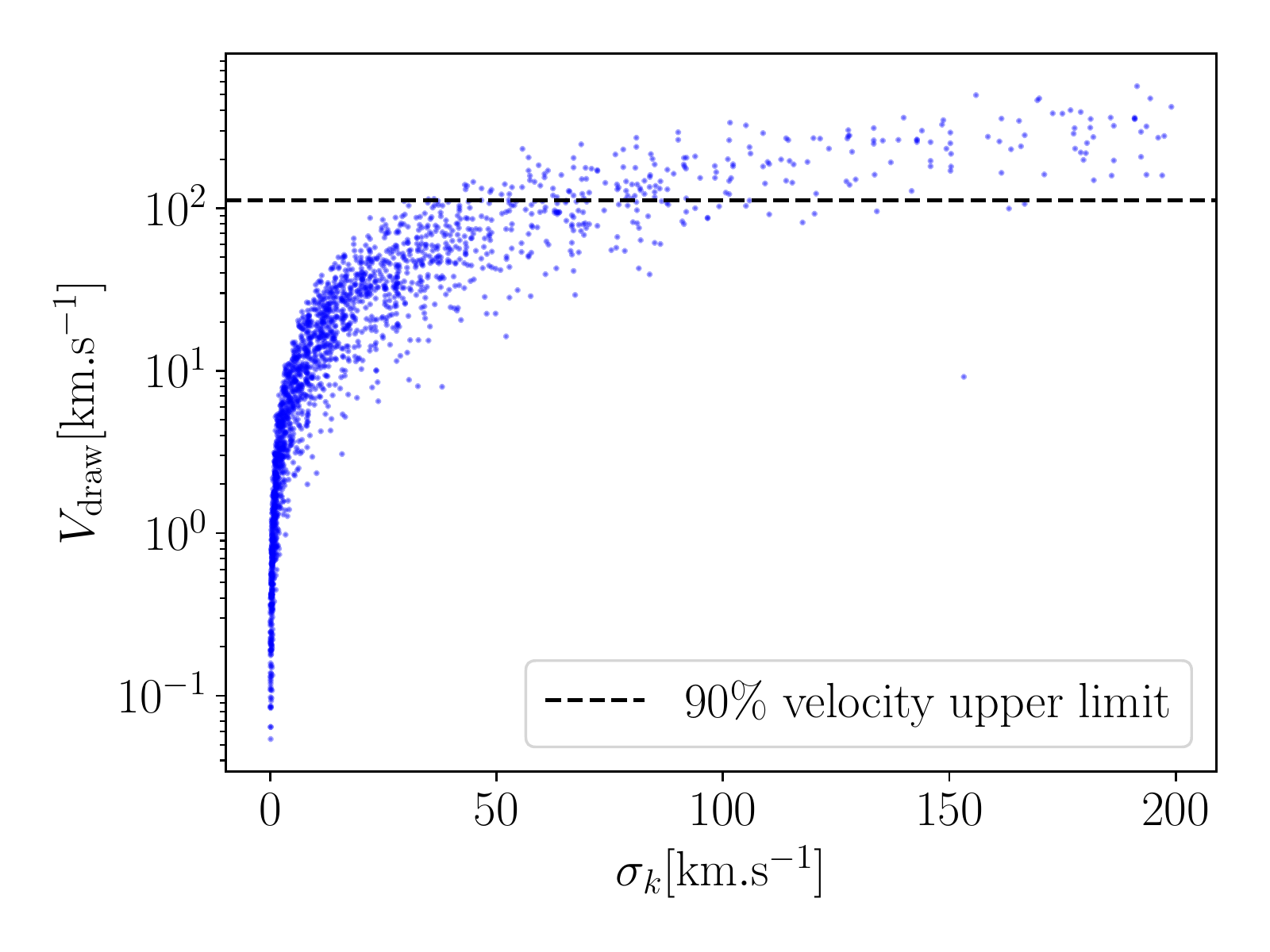}
    \caption{Kick velocity draws from the posterior on $\sigma_k$}
    \label{fig:sigma_draws}
\end{figure}

Assuming that natal kicks are drawn from the same distribution for all BHs, we can ask what their distribution might be like given the constraints on $\sigma_k$ from VFTS 243. Figure~\ref{fig:sigma_draws} shows 1000 draws of the velocity from the posterior, using Eq.~\ref{Eq:Maxwell_prior}. We see that most of the velocities are somewhat small with a $90 \%$ upper limit of \redtext{$110$} \kms. This is in line with most other of the kick measurements, for example from~\cite{Willems:2004kk,  Wong:2011eg, Wong:2013vya, Atri:2019fbx, Sanchez:2021lud} and other potential measurements~\citep{Andrews:2022cwi}, even in cases where a non-zero kick velocity is measured such as~\citet{Fragos:2008hg} and \citet{Kimball:2022xbp}.

Understanding the distribution of kick velocities has important implications for compact binary formation. Large kick velocities can unbind systems that form in the in-field channel. Even in dynamical formation channels like a globular cluster, large kicks can eject a BH out(see for example~\cite{Antonini:2016gqe, Fragione:2021nhb, Gerosa:2021mno}). A precise measurement of the kick distribution is thereby one of the factors that influence the calculation of merger rates from different binary formation channels, and the formation of higher mass black holes through repeated mergers.

\section{conclusion}
\label{Sec:conclusion}
In this paper, we have developed a statistical framework for inferring black hole natal kicks from binaries with a black hole and a detached luminous companion, and we have applied it to the system VFTS 243. We find that the black hole in VFTS 243 received a kick less than $72$ \kms, assuming a Maxwellian prior over the kick velocity magnitude and a broad uninformative mass prior. This finding is in line with a previous estimate from \cite{Stevance:2022cqo}, but is more conservative, utilizing just the data and the orbital physics of a binary. This provides evidence that black hole natal kicks can be small compared to those of neutron stars. 

Binary synthesis models predict that a large number of wide mixed binaries with a BH, potentially detectable by Gaia, exist in our galaxy~\citep{Breivik:2017cmy, Chawla2022}. The techniques developed in this paper present an important step towards distribution of BH kicks and their dependence on remnant and progenitor masses, binary companion parameters and other details, helping us to ultimately understand the physics of supernova better for BH remnants. Moreover, we are already seeing hints that some of these binaries potentially could have had wide eccentric orbits before SN~\citep{El-Badry:2022}. Future work will therefore focus on generalizing the formalism developed here to relax the assumption that the pre-SN orbit was circular. 

We have made some fundamental assumptions in this work, which future work could loosen. Perhaps most importantly, we have assumed the pre-SN orbital eccentricity was negligible and that the post-SN orbital parameters are not affected by further binary interactions. Additionally, we have only included selection effects due to binaries becoming unbound from kicks, but have ignored observational selection effects, which will also become important as we get more astrometric binaries from Gaia. With the statistical framework we have developed herein, we will be prepared to analyze more of these BH and luminous companion binaries from Gaia and other missions, ultimately paving the way for a full understanding of how black holes get their natal kicks.


\section{Acknowledgements}
We thank Shenar Tomer and Leonardo A. Almeida for providing data from their analysis and useful discussion on the data products. S.B. was supported by National Science Foundation grant PHY-2207945. Z.D. acknowledges support from the CIERA Board of Visitors Research Professorship. V.K. was partially supported through a CIFAR Senior Fellowship, a Guggenheim Fellowship, the Gordon and Betty Moore Foundation (grant award GBMF8477), and from Northwestern University, including the Daniel I. Linzer Distinguished University Professorship fund. C.K acknowledges support from Northwestern University. This research was supported in part through the computational resources and staff contributions provided for the Quest high performance computing facility at Northwestern University which is jointly supported by the Office of the Provost, the Office for Research, and Northwestern University Information Technology. \\

\bibliography{paper}{}
\bibliographystyle{aasjournal}

\appendix

\section{Jointly constraining orbital parameters and systemic velocity}
\label{Sec:Appdx1}

Let the pre-SN CM radial velocity of the binary be $\Gamma_{i}$. The exact velocity is unknown, but drawn from a normal distribution with with $\mu_0 = 271.6 \text{ and } \sigma_0 = 12.2$ \kms from spectroscopic measurements of nearby B-type stars~\citep{EvansCJ:2015}.

\begin{equation}
    p(\Gamma_i) = \frac{1}{\sqrt{2 \pi \sigma^2_0}} \exp \left [ - \frac{(\Gamma_i - \mu_0 )^2}{2\sigma_0^2} \right]. 
    \label{Eq:vel_prior}
\end{equation}
The post-SN center of mass radial velocity is related to the pre-SN one as

\begin{equation}
    \Gamma_f = \Gamma_i + \delta V_{CM}^\iota. 
\end{equation}

Figure~\ref{Fig:Orbital_schematic} depicts the angle between the line of sight and the systemic velocity kick, $V_{CM}$. Clearly, $\delta V_{CM}^{\iota} = \delta V_{CM} \cos \iota_{v}$. The angle $\iota_{v}$ physically has no bearing on the kick direction and neither should it depend on the pre-SN CM velocity of the system. A natural prior on it is uniformly isotropic. Now to use the joint constraints from the systemic velocity and the orbital parameters, we first marginalize over the kick directions, $\iota_v$ and $\Gamma_{i}$:~\footnote{We assert even though the CM velocity is completely determined by the kick velocity, masses and orbital parameters, $\iota_v$ is an independent angle that is more akin to the inclination angle, since it to the line of sight. }

\begin{equation}
    \small
    \pi (\alpha', e_f', \Gamma_f'  | V_k) = \int  d \left(\cos \theta\right)  \, d \phi \,  d \Gamma_i d (\cos \iota_v) \, \pi(\alpha', e_f' , \Gamma_f' | \phi, \cos \theta, \Gamma_i, V_k, \cos \iota_v) \, \pi(\phi) \, \pi(\cos\theta) \, \pi(\Gamma_i) \,   \pi (\cos \iota_v).
\end{equation}
The angular priors are all isotropic, i.e. $\pi(\phi) \, \pi(\cos\theta) = 1/4 \pi$ and $\pi (\cos \iota_v) = 1/2$. All probabilities here also depend on the masses through $\beta$, but we suppress that explicitly for purpose of clarity. Now, consider the conditional probability distribution. It can be split up as,

\begin{equation}
    \pi(\alpha', e_f' , \Gamma_f' | \phi, \cos \theta, \Gamma_i, V_k, \cos \iota_v) = \pi(\alpha', e_f' |\phi, \cos \theta,  V_k) \times  \pi( \Gamma_f' | \alpha', e_f',\phi, \cos \theta, \Gamma_i, V_k, \cos \iota_v), 
\end{equation}
where the orbital parameters, $\alpha, e$ are deterministically known for a given kick direction and velocity.  
\begin{equation}
\begin{split}
    \alpha = & \,  \frac{\beta}{2 \beta - v^2_x - (v_y + 1)^2 - v^2_z} \\
    1 - e^2 =  & \, ((v_y + 1)^2 + v^2_z) \, \frac{ (2 \beta - v^2_x - (v_y + 1)^2 - v^2_z) }{\beta^2}
\end{split}
\end{equation}
Similarly $\Gamma_f$ is also deterministically known,

\begin{equation}
    \Gamma_f = \Gamma_i + \delta V_{CM} \cos \iota_{v}. 
\end{equation}
Therefore the distribution of these quantities is given by delta functions, 
\begin{equation}
    \pi(\alpha', e_f' , \Gamma_f' | \phi, \cos \theta, \Gamma_i, V_k, \cos \iota_v) = \delta(\alpha - \alpha') \delta(e_f - e_f') \delta(\Gamma_f - \Gamma_f'). 
    \label{Eq:CM_vel_transform}
\end{equation}

We now perform a transformation of variables $\left \{ \phi, \cos \theta, \Gamma_i \right \} \longrightarrow \left \{\alpha, e_f, \Gamma_f \right \}$. The Jacobian is given by, 

\begin{equation}
\mathcal{J}(\alpha, e_f, \Gamma_f ) =   
\begin{Vmatrix}
\frac{\partial \phi}{\partial \alpha} & \frac{\partial \phi}{\partial e_f} & \frac{\partial \phi}{\partial \Gamma_f}\\
\frac{\partial \cos \theta}{\partial \alpha} & \frac{\partial \cos \theta}{\partial e_f} & \frac{\partial \cos \theta}{\partial \Gamma_f} \\ 
\frac{\partial  \Gamma_i}{\partial \alpha} & \frac{\partial \Gamma_i}{\partial e_f} & \frac{\partial \Gamma_i}{\partial \Gamma_f}
\end{Vmatrix}.
\end{equation}

\noindent But physically, the only thing that can depend on the initial CM velocity is the final velocity, so that $\frac{\partial \cos \Gamma_i}{\partial \alpha} = 0$ and $\frac{\partial \cos \Gamma_i}{\partial e_f} = 0$. We also see that $ \frac{\partial  \Gamma_i}{\partial \Gamma_f} = 1$ from Eq.~\ref{Eq:CM_vel_transform}. This implies, 

\begin{equation}
\mathcal{J}(\alpha, e_f, \Gamma_f ) =  \begin{vmatrix} \frac{\partial  \Gamma_i}{\partial \Gamma_f} \end{vmatrix} 
\begin{Vmatrix}
\frac{\partial \phi}{\partial \alpha} & \frac{\partial \phi}{\partial e_f} \\
\frac{\partial \cos \theta}{\partial \alpha} & \frac{\partial \cos \theta}{\partial e_f} 
\end{Vmatrix}. 
\end{equation}
This $2 \times 2$ determinant has been solved by ~\citet{Andrews:2019ome}.  Therefore we get, 

\begin{equation}
    \mathcal{J}(\alpha, e_f, \Gamma_f ) = \frac{4 e V_r}{\pi \alpha V_k} \sqrt{\frac{\beta^3}{C_1 (\alpha, e) \, C_2 (\alpha, e)}} . 
\end{equation}

Putting everything together, 
\begin{equation}
    \small
    \pi (\alpha', e_f', \Gamma_f'  | V_k) = \frac{V_r \beta^{3/2}}{2 \pi^2}\int d \alpha \, d e_f \, d \Gamma_f \, d (\cos \iota_v)\frac{e \, \pi(\Gamma_i)}{ \alpha V_k \sqrt{C_1 (\alpha, e) C_2 (\alpha, e) }} \delta(\alpha - \alpha') \delta(e_f - e_f') \delta(\Gamma_f - \Gamma_f'), 
\end{equation}
and working through the Dirac-delta functions, 

\begin{equation}
    \pi (\alpha', e_f', \Gamma_f'  | V_k) = \frac{V_r \beta^{3/2}}{2 \pi^2} \int_{-1}^{1} d (\cos \iota_v) \frac{e_f' \, \pi(\Gamma_f' - \delta V_{CM} \cos \iota_{v})}{ \alpha' V_k \sqrt{C_1 (\alpha', e_f') C_2 (\alpha', e_f') }} . 
\end{equation}
We now use Eq.~\ref{Eq:vel_prior} to get, 

\begin{equation}
    \pi (\alpha', e_f', \Gamma_f'  | V_k) = \frac{V_r e_f' \beta^{3/2} }{ 2 \pi^2\alpha' V_k \sqrt{C_1 C_2  }} \int_{\Gamma_f' - \delta V_{CM}}^{\Gamma_f + \delta V_{CM}} \frac{dt}{\sqrt{2 \pi \sigma^2_0}} \exp \left \{  - \frac{(t - \mu_0)^2}{2 \sigma^2_0}\right \}. 
\end{equation}
This finally yields, 

\begin{equation}
\small
    \pi (\alpha', e_f', \Gamma_f'  | V_k) = \frac{V_r e_f' \beta^{3/2} }{ (2 \pi)^{5/2} \sigma_0 \alpha' V_k \sqrt{C_1 C_2  }} \left \{ \erf \left(\frac{\Gamma_f + \delta V_{CM} - \mu_0}{\sqrt{2 \sigma_0^2}} \right) - \erf \left(\frac{\Gamma_f - \delta V_{CM} - \mu_0}{\sqrt{2 \sigma_0^2}} \right)   \right \}. 
\end{equation}

We still need to figure out what $\delta V_{CM}$ is, as a function of $\alpha, e_f$, since we have marginalized over the kick angles. Following~\citet{Andrews:2019ome} we have 

\begin{equation}
    \frac{V_{CM}^2}{V^2_r} = k_1 + k_2 \frac{2 \alpha - 1}{\alpha} - k_3 \frac{v_k \cos \theta + 1}{\sqrt{\beta}}. 
\end{equation}
Let's rewrite $\cos \theta$ in terms of $\alpha, \beta$ and $V_k$

\begin{equation}
    \cos \theta = \frac{\beta}{ 2 v_k} \left( 2  - \frac{1}{\alpha} \right) - \frac{v^2_k + 1}{ 2 v_k}. 
\end{equation}
Therefore, 

\begin{equation}
    \frac{V^2_{CM} (\alpha, \beta, V_k)}{V^2_r} = k_1 +   k_2 \, \frac{2 \alpha - 1}{\alpha} - \frac{k_3 \sqrt{\beta}}{2} \left( 2  - \frac{1}{\alpha} \right) + \frac{k_3}{2 \sqrt{\beta}} \left( v^2_k + 1 \right) - \frac{k_3}{\sqrt{\beta}}. 
\end{equation}



\end{document}